\documentclass[utf8]{frontiersFPHY} 
\setcitestyle{square} 
\usepackage{url,hyperref,lineno,microtype,subcaption}
\usepackage[onehalfspacing]{setspace}
\usepackage{graphicx,amssymb}
\usepackage{hyperref,amsmath,ulem,color}

\def\keyFont{\fontsize{8}{11}\helveticabold}
\def\firstAuthorLast{Gupta {et~al.}} 
\def\Authors{Shamik Gupta\,$^{1,2,*}$ and Arun M. Jayannavar\,$^{3,\dagger}$}
\def\Address{$^{1}$Department of Physics, Ramakrishna Mission Vivekananda Educational and Research Institute, Belur Math,  Howrah 711202, India \\ $^2$Regular Associate, Quantitative Life Sciences Section, ICTP–Abdus Salam International Centre for Theoretical Physics,
Strada Costiera 11, 34151 Trieste, Italy\\
$^3$Bhaurao Kakatkar College,  Jyoti Compound Camp,  Club Road Belgaum, India\\
$\dagger$Deceased}
\def\corrAuthor{Corresponding Author}
\def\corrEmail{shamikg1@gmail.com}

\begin{document}
\newcommand{\cor}[1]{\textcolor{red}{\bf{#1}}}
\onecolumn
\firstpage{1}

\title[Stochastic resetting: A (very) brief review]{Stochastic resetting: A (very) brief review} 

\author[\firstAuthorLast]{\Authors} 
\address{} 
\correspondance{} 

\extraAuth{}
\maketitle

\begin{abstract}
Stochastic processes offer a fundamentally different paradigm of dynamics than deterministic processes,  the most prominent example of the latter being Newton's laws of motion. Here, we discuss in a pedagogical manner a simple and illustrative example of stochastic processes in the form of a particle undergoing standard Brownian diffusion, with the additional feature of the particle resetting repeatedly and at random times to its initial condition. Over the years, many different variants of this simple setting have been studied,  all of which serve as illustrations of non-trivial and interesting static and dynamic features that characterize stochastic dynamics at long times. We will provide in this work a brief overview of this active and rapidly evolving field by considering the arguably simplest example of Brownian diffusion in one dimension. Along the way, we will learn about some of the general techniques that a physicist employs to study stochastic processes. Relevant to the special issue, we will discuss in detail how introducing resetting in an otherwise diffusive dynamics provides an explicit optimization of the time to locate a target through a special choice of the resetting protocol.  We also discuss thermodynamics of resetting, and provide a bird's eye view of some of the recent work in the field of resetting.
\tiny
 \keyFont{ \section{Keywords:} Stochastic processes, Stochastic resetting, Fokker-Planck equation,  Stationary state,  First-passage time}
\end{abstract}

\section{Introduction: Brownian motion}
Brownian diffusion models random motion of a mesoscopic particle that is immersed in a fluid and is being constantly buffeted by the fluid molecules. 
Starting from this simple context,  the paradigm of Brownian diffusion has been successfully employed to discuss a wide range of dynamical scenarios in physics, astronomy,  chemistry,  biology and mathematics, and even in finance and computer science. 
Consider a Brownian particle undergoing diffusion in free space in one dimension.  Its dynamics is conveniently described by the so-called overdamped Langevin equation for the displacement of its center of mass, and is given by \cite{Zwanzig}:  
\begin{align}
\frac{{\rm d}x}{{\rm d}t}=\eta(t),
\label{eq:eom}
\end{align}
where $\eta(t)$ is a Gaussian, white noise with the properties
\begin{align}
\langle \eta(t)\rangle=0,~~\langle \eta(t)\eta(t')\rangle=2D\delta(t-t').
\label{eq:noise-properties}
\end{align}
Here,  angular brackets denote average over noise realizations. The parameter $D>0$,  called the diffusion coefficient,  sets the strength of the noise.  

The dynamics~(\ref{eq:eom}) is an example of a Markov process, namely, a process that evolves in continuous time and for which if one wants to know at any instant of time the future state of the process,  it suffices to know the state of the process at that instant, and one does not need to know the entire history of the process until that instant.  Indeed, it is evident from Eq.~(\ref{eq:eom}) that to know at any instant of time $t$ the future state $x(t+\Delta t);~\Delta t >0$,  one requires to know just $x(t)$ and not the entire history from the initial state $x(t=0)$ to $x(t)$.  Note that the state space spanned by values of $x$ is continuous.

For a given initial condition $x(t=0)=x_0$,  Eq. (\ref{eq:eom}) may be integrated to obtain
\begin{align}
x(t)=x_0+\int_0^t {\rm d}t'\,\eta(t'),
\label{eq:xsoln}
\end{align}
which implies that for a given initial condition,  the position of the particle at time $t$ may have many different values depending on the trajectory of the noise $\eta$ between times $0$ and $t$.   Equation (\ref{eq:eom}) is an example of what are known as stochastic processes,  in which one may have many different final outcomes for the same initial condition.  This latter fact may be contrasted with what happens under deterministic processes,  the most prominent example of which would perhaps be the Newton's laws of motion, which have the property of yielding a unique final outcome for a given initial condition.  Equation (\ref{eq:eom}) is more specifically known as a stochastic differential equation: a differential equation in which one or more terms is a stochastic process,  and whose solution is consequently also a stochastic process. 

The average position of the particle at time $t$, averaged over the different noise trajectories, is obtained from Eq.  (\ref{eq:xsoln}) on using Eq.  (\ref{eq:noise-properties}) as $\langle x(t)\rangle=x_0$, while the variance, measured with respect to the initial location of the particle, reads
\begin{align}
\langle( x(t)-x_0)^2\rangle(t)=\int_0^t \int_0^t {\rm d}t_1 {\rm d}t_2 \, \langle \eta(t_1)\eta(t_2)\rangle =2Dt.
\label{eq:xvariance}
\end{align}

\subsection{Position probability distribution}
Now,  it is instructive to ask: given that the position of the particle at time $t$  has many different values, what is the full distribution of the position at time $t$? In other words, what is the form of $P(x,t)$,  defined such that $P(x,t){\rm d}x$ is the probability to find the particle between positions $x$ and $x+{\rm d}x$ at time $t$, given that the particle started from position $x_0$ at time $t=0$: $P(x,0)=\delta(x-x_0)$.  The probability density is of course normalized as $\int {\rm d}x\, P(x,t)=1\,\forall\, t$.  The question just posed may be answered by writing down and solving the so-called Fokker-Planck equation for the time evolution of $P(x,t)$. The equation reads
\begin{align}
\frac{\partial P(x,t)}{\partial t}=D\frac{\partial^2 P(x,t)}{\partial x^2},
\label{eq:FPE}
\end{align}
and needs to be solved subject to the initial condition $P(x,0)=\delta(x-x_0)$. 

The Fokker-Planck equation may be derived by noting down the ways in which the probability density $P(x,t)$ changes in a small time $0 < \Delta t \ll 1$, and finally taking the limit $\Delta t \to 0$.  One has 
\begin{align}
P(x,t+\Delta t)=P(x,t)+\int {\rm d}(\Delta x)\,P(x-\Delta x,t)\phi_{\Delta t}(\Delta x)-\int {\rm d}(\Delta x) \, P(x,t)\phi_{\Delta t}(\Delta x),
\label{eq:FPE-derivation}
\end{align}
where $\phi_{\Delta t}(\Delta x)$ is the probability density for $x$ to change by an amount $\Delta x$ in time $\Delta t$,  and is taken to be independent of $x$.  The probability density satisfies the normalization $\int {\rm d}(\Delta x)\, \phi_{\Delta t}(\Delta x)=1$.  Note that Eq.  (\ref{eq:eom}) implies that this jump probability is independent of the value of $x$ from which the jump is taking place.  In Eq. (\ref{eq:FPE-derivation}), the second term on the right hand side (rhs) denotes gain in probability due to a change taking place from $x-\Delta x$ to $x$, while the third term on the rhs denotes loss in probability due to a change taking place from $x$.   Next,  noting that the dynamics (\ref{eq:eom}) implies that for small $\Delta t$, the change $\Delta x$ that $x$ undergoes is also small, and assuming that $P(x,t)$ is a slowly varying function of $x$,  we may Taylor expand the left hand side (lhs) of Eq. (\ref{eq:FPE-derivation}) in powers of $\Delta t$ and the second term on the right hand side in powers of $\Delta x$.  Executing such an expansion, and noting that Eq. (\ref{eq:eom}) implies that $\langle \Delta x \rangle \equiv \int {\rm d}(\Delta x)\,\Delta x\, \phi_{\rm \Delta t}(\Delta x)=0$ and $\langle (\Delta x)^2 \rangle \equiv  \int {\rm d}(\Delta x)\,(\Delta x)^2\, \phi_{\rm \Delta t}(\Delta x)=2D\Delta t$, one obtains Eq.  (\ref{eq:FPE}).

\begin{figure}
\centering
\includegraphics[width=85mm]{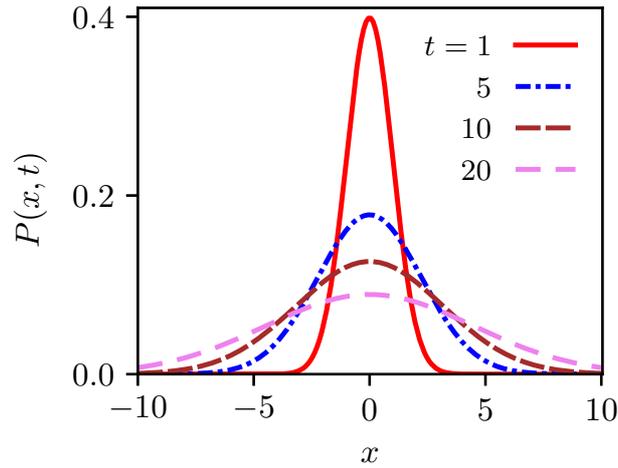}
\caption{The distribution (\ref{eq:xdistribution}) at different times, showing the broadening of the distribution with time. Here, we have taken $x_0=0.0,D=0.5$.}
\label{fig:Pxt-Gaussian}
\end{figure}

The solution to Eq.  (\ref{eq:FPE}) is readily obtained by expressing $P(x,t)$ in terms of its Fourier component $\widetilde{P}(k,t)=\int_{-\infty}^\infty {\rm d}x \, P(x,t)\exp(-{\rm i}kx)$ that satisfies a linear differential equation given by $\partial \widetilde{P}(k,t)/\partial t=-Dk^2\widetilde{P}(k,t)$ subject to the initial condition $\widetilde{P}(k,0)=\exp(-{\rm i}kx_0)$; this equation has the solution $\widetilde{P}(k,t)=\exp(-Dk^2 t -{\rm i}kx_0)$.  On performing inverse Fourier transformation  according to the prescription $P(x,t)=1/(2\pi)\int_{-\infty}^\infty {\rm d}k\, \widetilde{P}(k,t)\exp({\rm i}kx)$,  one finally obtains
\begin{align}
P(x,t)=\frac{1}{\sqrt{4\pi Dt}}e^{-(x-x_0)^2/(4Dt)},
\label{eq:xdistribution}
\end{align}
which is a Gaussian distribution centred at $x_0$.  Equation (\ref{eq:xdistribution}) implies that the probability of finding the particle at a given distance from its initial location grows with time. Consequently, as time goes by,  one has an increased probability of finding the particle far off from its initial location, although at any given time, the most probable location of the particle is at its initial location.  These features are evident from the plot of $P(x,t)$ at different times depicted in Fig.  \ref{fig:Pxt-Gaussian}. 

The mean-squared displacement (MSD) of the particle about its initial location is obtained straightforwardly from Eq.~(\ref{eq:xdistribution}) as
\begin{align}
\langle(x-x_0)^2\rangle(t)=\int_{-\infty}^\infty {\rm d}x~(x-x_0)^2P(x,t)=2Dt,
\label{eq:MSD-BM}
\end{align}
matching with the result~(\ref{eq:xvariance}), and 
which implies that the MSD grows forever as a function of time. Such a behavior in which the MSD grows linearly in time is referred to as normal diffusive behavior. 

The aspect of the probability of finding the particle far off from $x_0$ increasing with time may be traced back to the fact that since the particle is diffusing in free space, it is no wonder that at longer times, it would have spread to a larger region of the available space than at shorter times.  Let us then ask this question: Is there a way to stop this spread?  Well, one possible way could be to have the particle diffuse not in free space but in a bounded domain of finite extent, say, between points $x=-L$ and $x=+L$ on the $x$-axis.  In this case,  the probability of finding the particle at a given location would keep increasing with time, until it can increase no further, that is,  until the system attains a stationary state. In this state,  attained in the limit $t \to \infty$,  it may be argued that this probability would be the same at all locations inside the bounded domain. In other words, the particle would be equally likely to be found anywhere within the region $x \in [-L,L]$. 

\begin{figure}
\centering
\includegraphics[width=85mm]{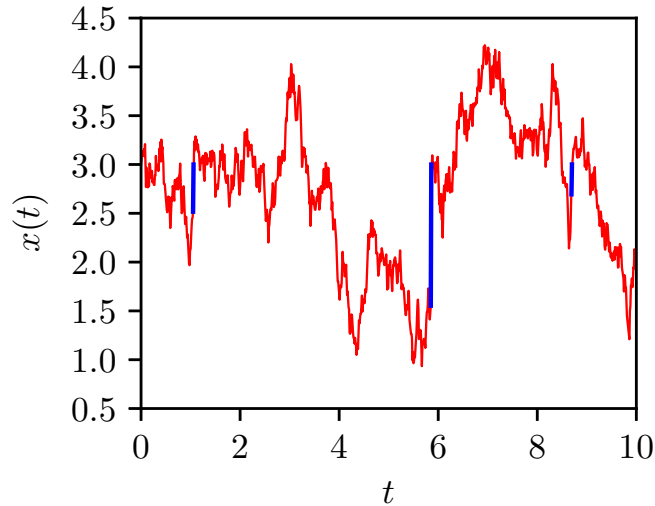}
\caption{The plot shows a typical trajectory of a particle starting at location $x_0=3.0$ and performing free diffusion in one dimension, with stochastic resets to $x_0$ at random times.  The parameter values are $D=0.5$, and $r=0.25$. The jumps in $x$ values corresponding to resets are marked in blue.} 
\label{fig:trajectory}
\end{figure}

\section{Brownian motion in presence of resetting}
In the above backdrop, one may ask: Is there a way to induce a stationary state in the system by not tweaking the boundary conditions but by modifying the dynamics in a way that it continues to take place in the free space and yet has a stationary state ? Of course, one way could be to subject the motion of the particle to take place in a bounded potential.  In this respect,  tweaking the boundary condition so that the particle diffuses not in free space but in a bounded domain of finite extent is tantamount to a potential that is zero everywhere except at the boundaries where one has an infinitely-high potential barrier.  As an interesting alternative to using a potential,  it turns out that there is a rather simple and instructive way to achieve the goal of inducing a stationary state, through the introduction of stochastic resetting in the dynamics~\cite{Evans:2011}.  To this end, let us modify the dynamics of free diffusion by stipulating that the particle in addition to evolving according to the rule (\ref{eq:eom}) has also the option of resetting its position to its initial value. More specifically,  at time $t$, the particle has in the ensuing infinitesimal time interval ${\rm d}t$ the following options for updating its location $x(t)$: with probability $r{\rm d}t$, it resets its position to initial value $x_0$, while with the complementary probability $1-r{\rm d}t$, it evolves according to Eq. (\ref{eq:eom}). Thus, we have \cite{Evans:2011}
\begin{align}
x(t+{\rm d}t)=\left\{ 
\begin{array}{ll}
               x_0 & \mbox{with\,prob.\, $r{\rm d}t$}, \\
               x(t)+\eta(t){\rm d}t & \mbox{with \, prob.\, $1-r{\rm d}t$}.
               \end{array}
        \right. 
        \label{eq:resetting-rules}
\end{align}
Here, $r\ge0$ is a dynamical parameter,  whose value when set to zero reduces the modified dynamics to that of free diffusion. In practical terms, a typical trajectory of the particle would involve  free diffusion interspersed with events of reset to $x_0$, as shown in Fig. \ref{fig:trajectory}.  It is evident from Eq.~(\ref{eq:resetting-rules}) that introducing resetting into the dynamics~(\ref{eq:eom}) retains the Markov nature of the dynamics.

\subsection{Position probability distribution}
Suppose we ask: what is the probability for the next reset to happen after a certain time $t$, in the interval $[t,t+{\mathrm d}t]$? Let us discretize time in equal steps of length $0 < \Delta t \ll 1$; we will eventually consider the limit $\Delta t \to 0$.  The number of such steps during time duration $t$ is obviously given by ${\mathcal N}=t/\Delta t$.  During each step of length $\Delta t$,  the probability of reset equals $r\Delta t$, while the same for no reset equals $1-r\Delta t$.  Consequently,  the probability that the next reset happens after time $t$, in the interval $[t,t+\Delta t]$,  is given by $\underbrace{(1-r\Delta t)(1-r\Delta t)\ldots(1-r\Delta t)}_{{\mathcal N}\, \mathrm{factors}}r\Delta t=(1-r\Delta t)^{\mathcal N}r\Delta t=(1-r\Delta t)^{t/\Delta t}r\Delta t$.  In the limit $\Delta t \to 0$, the latter quantity equals $\exp(-rt)\,r {\rm d}t$.  We thus conclude that the probability for no reset to occur during a given time duration $t$ equals $\exp(-rt)$, while the probability for the next reset to occur after time $t$, that is, in the interval $[t,t+{\rm d}t]$, equals $\exp(-rt)\,r{\rm d}t$.  For latter purpose,  let us note that at a given time instant $t$,  the probability that the last reset happened in the interval $[t-\tau-{\rm d}\tau, t-\tau]$, with $\tau \in [0,t]$, equals $\exp(-r\tau)\,r{\rm d}\tau$.  

Note that as mentioned earlier,  the dynamics (\ref{eq:eom}) represents infinitesimal change in $x$ in infinitesimal time interval ${\rm d}t$, while incorporating resetting into the dynamics has in the resulting dynamics reset events leading to any amount of change in $x$ in infinitesimal interval ${\rm d}t$. Indeed, the reset destination is always the same,  namely, $x_0$, irrespective of the location the particle resets from,  so that the particle may execute a very long jump during a reset event. 

 It should be apparent from the aforementioned dynamical rules that introducing resetting into the dynamics would not let the particle go at any time, long or short, very far from the initial location.  Is it actually the case? To answer this, we need to study just as in the case of free diffusion discussed above the time evolution of the probability density $P_{\rm r}(x,t)$ in presence of resetting, which may be derived by writing down the ways in which $P_{\rm r}(x,t)$ changes in a small time $0<\Delta t\ll 1$ and then taking the limit $\Delta t \to 0$.  One has
\begin{align}
P_{\rm r}(x,t+\Delta t)&=
P_{\rm r}(x,t)+(1-r\Delta t)\left[\int {\rm d}(\Delta x)\,P_{\rm r}(x-\Delta x,t)\phi_{\Delta t}(\Delta x)-\int {\rm d}(\Delta x) \, P_{\rm r}(x,t)\phi_{\Delta t}(\Delta x) \right]\nonumber \\
&-r\Delta t P_{\rm r}(x,t)+r\Delta t \delta(x-x_0).
\label{eq:FPE-reset-derivation}
\end{align}
Here,  the second and the third term arise from diffusion, while the fourth and the fifth term are owing to resetting events.  While the former two terms are as in the free diffusion case, Eq. (\ref{eq:FPE-derivation}),  the latter two terms may be understood as follows. 
For $r\ne 0$, there is a loss in probability at all locations other than $x_0$ due to resetting events, which is represented by the fourth term on the rhs of Eq.  (\ref{eq:FPE-reset-derivation}), while the corresponding gain in probability at $x=x_0$, given by $r\Delta t\delta(x-x_0)\int {\rm d}x \, P_{\rm r}(x,t)=r\Delta t\delta(x-x_0)$,  yields the last term on the rhs.  Here we have used the normalization condition $\int {\rm d}x\, P_{\rm r}(x,t)=1\,\forall\,t$. 
Implementing appropriate Taylor expansions on both sides of Eq. (\ref{eq:FPE-reset-derivation}),  then taking the limit $\Delta t \to 0$, the probability $P_{\rm r}(x,t)$ may be shown to be satisfying the Fokker-Planck equation~\cite{Evans:2011}
 \begin{align}
 \frac{\partial P_{\rm r}(x,t)}{\partial t}=D\frac{\partial^2 P_{\rm r}(x,t)}{\partial x^2}-rP_{\rm r}(x,t)+r\delta(x-x_0).
\label{eq:reset-FPE} 
 \end{align}
The initial condition is 
 $P_{\rm r}(x,0)=\delta(x-x_0)$.  Equation (\ref{eq:reset-FPE}) may be solved for $P_{\rm r}(x,t)$; here, we report only the solution, referring the reader to Ref. \cite{Giuggioli:2019} for details of obtaining it:
\begin{align}
P_{\rm r}(x,t)&=\frac{e^{-rt-\frac{(x-x_0)^2}{4Dt}}}{\sqrt{4\pi Dt}}+\frac{1}{4}\sqrt{\frac{r}{D}}e^{-\sqrt{\frac{r}{D}}|x-x_0|}{\rm Erfc}\left(\frac{|x-x_0|}{2\sqrt{Dt}}-\sqrt{rt}\right) \nonumber \\
& -\frac{1}{4}\sqrt{\frac{r}{D}}e^{\sqrt{\frac{r}{D}}|x-x_0|}{\rm Erfc}\left(\frac{|x-x_0|}{2\sqrt{D
t}}+\sqrt{rt}\right),
\label{eq:reset-Pxt}
\end{align}
where ${\rm Erfc}(x) \equiv 2/\left(\sqrt{\pi}\right) \int_x^\infty {\rm d}y \, e^{-y^2} $ is the complementary error function.  The MSD of the particle may be obtained from Eq.~(\ref{eq:reset-Pxt}) as
\begin{align}
\langle (x-x_0)^2\rangle(t)=\frac{2 D}{r}\left(1-e^{-rt}\right).
\end{align}

Equation (\ref{eq:reset-Pxt}) implies that in the limit $t\to \infty$, one has a stationary-state form \cite{Evans:2011}: 
\begin{align}
P_{\rm r, ss}(x)=\frac{1}{2}\sqrt{\frac{r}{D}}e^{-|x-x_0|\sqrt{r/D}}.
\label{eq:steadystate}
\end{align}
Correspondingly, the MSD relaxes at long times to the stationary-state value
\begin{align}
\langle (x-x_0)^2\rangle_{\rm r,ss}=\frac{2D}{r}.
\label{eq:MSD-BM-reset}
\end{align}

\begin{figure}
\centering
\includegraphics[width=85mm]{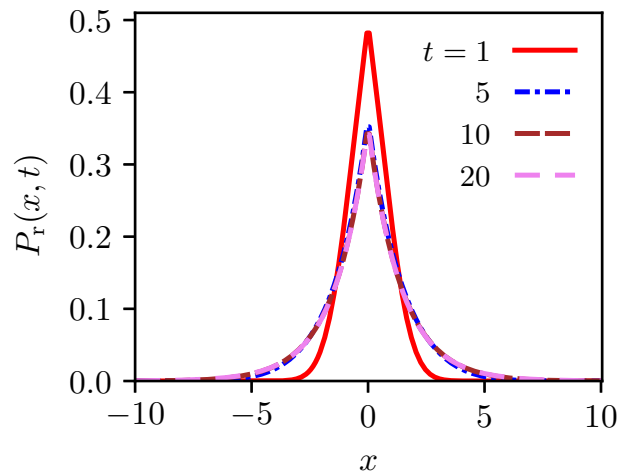}
\caption{Probability distribution $P_{\rm r}(x,t)$ in presence of resetting and at four different times,  as given by Eq.  (\ref{eq:reset-Pxt}),  showing convergence in time to a stationary state in which the distribution no longer changes with time. The latter is given by Eq. (\ref{eq:steadystate}).  The parameter values are $x_0=0.0$, $D=0.5$, and $r=0.25$.}
\label{fig:Pxt-resetting}
\end{figure}

Figure~\ref{fig:Pxt-resetting} shows the probability distribution $P_{\rm r}(x, t)$ in presence of resetting and at four different times, as given by Eq. ~(\ref{eq:reset-Pxt}), showing in particular convergence in time to the stationary state~(\ref{eq:steadystate}).

The stationary-state distribution is an exponential centred at the resetting location; since the particle keeps resetting to $x_0$ in time,  it is no wonder that the most likely position of the particle is $x_0$.  More importantly,  an exponential profile implies that there is a characteristic length scale within which the particle is to be found with significant probability and beyond which one has an exponentially-small probability of finding the particle.   In conclusion, we have been able to achieve our goal: by modifying not the boundary conditions but rather the dynamics of free diffusion, one induces a stationary state in the system.  While we have discussed in the foregoing the case of resetting at exponentially-distributed random times,  the case of resetting at power-law-distributed random times has also been considered in the literature, see Ref.~\cite{ref2}.  We remark in passing that resetting creates a source of probability at $x_0$ and a sink at all other locations $x \ne x_0$, so that the condition of detailed balance, which characterizes an equilibrium stationary state~\cite{Bertin-book}, is manifestly violated in the stationary state induced by resetting.  Consequently, the stationary state~(\ref{eq:steadystate}) is a generic nonequilibrium stationary state~\cite{Evans:2011}. 

\section{First-passage time distribution}
Now that we have seen a stationary state emerging in presence of stochastic resetting,  one may wonder if besides this issue of theoretical relevance there is any practical utility of the process of stochastic resetting.  Here, we will discuss one very interesting application of stochastic resetting, namely,  in the context of search processes. 

\begin{figure}
\centering
\includegraphics[width=85mm]{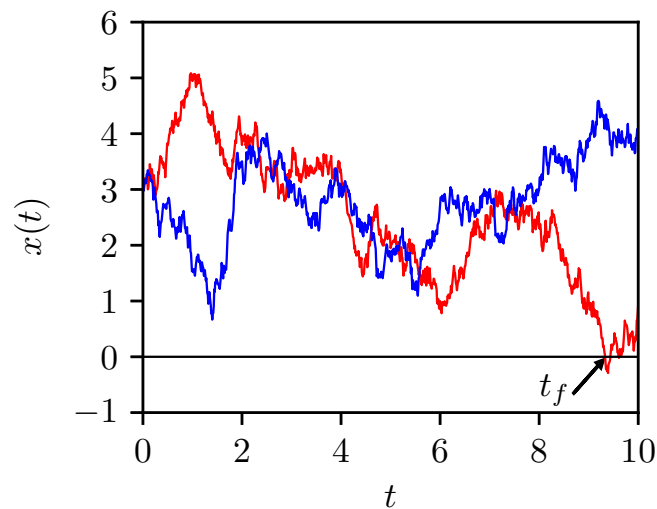}
\caption{Two sample trajectories $x(t)$ corresponding to free diffusion (\ref{eq:eom}) with initial condition $x(0)=x_0=3.0$, and with $D=0.5$. Here, the quantity $t_f$ is the first-passage time for the red trajectory.}
\label{fig:fpt}
\end{figure}

\subsection{The case of Brownian motion}
 In order to discuss the concept and utility of first-passage times,  let us come back to free diffusion.  Referring to Fig.  \ref{fig:fpt},  suppose we ask: when does a trajectory of free diffusion starting at $x_0>0$ at time $t=0$ cross the origin for the first time? Clearly the time $t_f$ that it happens, aptly termed the first-passage time (FPT),  is a random variable that varies from trajectory to trajectory, and one may ask: what is the form of its distribution ${\mathcal P}(x_0, t_f)$ ? This distribution would explicitly depend on $x_0$,  a fact encoded in the defining symbol for the distribution.  In order to answer this question,  noting that $t_f$ is by definition a positive quantity,  it proves convenient to consider the Laplace transform of ${\mathcal P}(x_0, t_f)$:
\begin{align}
\widetilde{\mathcal P}(x_0, s)\equiv \int_0^\infty {\rm d}t_f\,e^{-st_f}{\mathcal P}(x_0, t_f)=\langle e^{-st_f}\rangle,
\label{eq:Q}
\end{align}
where the angular brackets denote averaging with respect to the distribution ${\mathcal P}(x_0, t_f)$,  $s$ is the Laplace variable, and tilde denotes the function obtained on performing Laplace transform ${\mathcal L}$ of a function $f(t)$, that is, ${\mathcal L}[f(t)]=\widetilde{f}(s)$.  Now, let us understand that a given value of $t_f$ corresponds to all trajectories $\{x(\tau)\}_{0\le \tau \le t_f}$ between times $0$ and $t_f$ satisfying $x(0)=x_0$ and $x(0 < \tau < t_f)>0$.  Consequently,  averaging with respect to $t_f$ is tantamount to averaging with respect to all trajectories that start at $x_0$ and propagate for time $t_f$ without ever crossing the origin, and such an averaging is what the angular brackets in Eq.  (\ref{eq:Q}) would also stand for.
 
In order to proceed,  we now derive a differential equation for $\widetilde{\mathcal P}(x_0, s)$.  To this end,  following Ref.  \cite{Majumdar-rev},  we split the time interval $\lbrack 0, t_f \rbrack$ into two parts.  During the first interval $\lbrack 0,\Delta \tau \rbrack$,  
the aforementioned set of trajectories start from $x_0$ and propagate up to $x_0+\Delta x$,  where $\Delta x$ is of course different for different trajectories. 
In the second interval $\lbrack \Delta \tau, t_f\rbrack$,  these trajectories start at $x_0+\Delta x$ and reach $0$ at $t_f$.  We take $\Delta \tau$ to be a small time interval with $0<\Delta \tau \ll 1$, and eventually, we would take the limit $\Delta \tau \to 0$. 
We get, to leading order in $\Delta \tau$,  that
\begin{align}
\widetilde{\mathcal P}(x_0,s)= e^{-s\Delta \tau}\left\langle \widetilde{\mathcal P}(x_0+\Delta x,s)\right\rangle_{\Delta x}\approx (1-s\Delta \tau)\left\langle \widetilde{\mathcal P}(x_0+\Delta x,s)\right\rangle_{\Delta x},
\label{eq:Qint}
\end{align}
where the average denoted by the angular brackets is to be performed over all realizations of $\Delta x$.  Now, $\Delta \tau$ being small implies that so is $\Delta x$, so that the quantity $\widetilde{\mathcal P}(x_0+\Delta x,s)$ on the rhs of the above equation may be expanded in a Taylor series in $\Delta x$.  One gets
\begin{align}
\widetilde{\mathcal P}(x_0,s) \approx (1-s\Delta \tau)\left\langle \widetilde{\mathcal P}(x_0,s)+\widetilde{\mathcal P}'(x_0,s)\Delta x+\widetilde{\mathcal P}''(x_0,s)\frac{(\Delta x)^2}{2}+\ldots \right\rangle_{\Delta x}.
\label{eq:Qeqn}
\end{align}
Here,  the prime denotes derivative with respect to $x_0$. 
Now, Eq.  (\ref{eq:eom}) gives $\Delta x=\eta(0)\Delta \tau$, so that averaging with respect to $\Delta x$ is equivalent to averaging with respect to different realizations of $\eta(0)$. Using $\langle \eta(0)\rangle=0$ and $\langle \eta^2(0)\rangle=2D/\Delta \tau$ (see Appendix) so that $\langle \Delta x\rangle_{\Delta x}=0$ and $\langle (\Delta x)^2 \rangle_{\Delta x}=2D\Delta \tau$,  we get from Eq.  (\ref{eq:Qeqn}) on taking the limit $\Delta \tau \to 0$ that 
\begin{align}
D\widetilde{\mathcal P}''(x_0,s)-s\widetilde{\mathcal P}(x_0,s)=0.
\label{eq:Q-DE}
\end{align}
The above equation, referred to as the backward Fokker-Planck equation,  is to be solved subject to two boundary conditions: (i) $\widetilde{\mathcal P}(x_0 \to 0,s)=1,$ since as $x_0\to 0$,  we have $t_f \to 0$,  and (ii) $\widetilde{\mathcal P}(x_0\to \infty,s)=0$,  since as $x_0 \to \infty$, we have $t_f \to \infty$.  The solution of Eq.  (\ref{eq:Q-DE}) subject to these boundary conditions is
\begin{align}
\widetilde{\mathcal P}(x_0,s)=e^{-\sqrt{s/D}x_0}.
\end{align}
On performing inverse Laplace transformation, one readily gets \cite{Majumdar-rev}
\begin{align}
{\mathcal P}(x_0,t_f)=\frac{x_0}{\sqrt{4\pi D}}\frac{e^{-x_0^2/(4Dt_f)}}{t_f^{3/2}}.
\label{eq:PFPT-diffusion}
\end{align}
(For $x_0<0$,  the corresponding FPT distribution is obtained from the above equation with the replacement $x_0 \to |x_0|$. )
We thus find that the FPT distribution has a tail $\sim t_f^{-3/2}$, and moreover,  that the average FPT (the mean first-passage time (MFPT)) given by $\langle t_f \rangle = \int_0^\infty {\rm d}t_f~t_f P(x_0,t_f)$ is infinite ! The latter result implies that the dynamics (\ref{eq:eom}) allows for trajectories which starting at $x_0$ start to move away from the origin and hence take a very long time to hit the origin for the first time (e.g.,  the blue trajectory in Fig.  \ref{fig:fpt}).  Does repeated resetting to $x_0$ kill these trajectories,  given that it allows the particle to remain effectively in a finite region around $x_0$? Let us then investigate the distribution of the FPT in presence of resetting.

\subsection{The case of Brownian motion in presence of resetting}
In order to proceed, define ${\mathcal P}_{\rm r}(x_0,t_f)$ as the FPT distribution in presence of resetting, and $S_{\rm r}(x_0,t)$ as the so-called survival probability, that is, the probability that a trajectory starting at $x_0$ at time $t=0$ has not crossed the origin up to time $t$.  Note that resetting is taking place not to the origin but to the initial location $x_0$ of the particle.  Obviously then one has $S_{\rm r}(x_0,t+\Delta t)-S_{\rm r}(x_0,t)={\mathcal P}_{\rm r}(x_0,t)\Delta t$ for small $\Delta t$, which in the limit $\Delta t \to 0$ gives ${\mathcal P}_{\rm r}(x_0,t)=-\partial S_{\rm r}(x_0,t)/\partial t$. In terms of $S_{\rm r}(x_0,t)$,  the mean FPT is given as $\langle t_f \rangle_{\rm r} \equiv \int_0^\infty {\rm d}t_f \, t_f {\mathcal P}_{\rm r}(x_0,t_f)=-\int_0^\infty {\rm d}t_f\, t_f\partial S_{\rm r}(x_0,t_f)/\partial t_f=-[t_f S_{\rm r}(x_0,t_f)]_0^\infty+\int_0^\infty {\rm d}t_f \,S_{\rm r}(x_0,t_f)=\int_0^\infty {\rm d}t_f\,S_{\rm r}(x_0,t_f)= \widetilde{S}_{\rm r}(x_0,s=0)$, using $S_{\rm r}(x_0,\infty)=0$.   
A trajectory starting from $x_0$ that reaches the origin for the first time at time $t$ may have had no reset at all since the time instant $t=0$ or may have had its last reset at an earlier time instant in the interval $[t-\tau-{\rm d}\tau,t-\tau]$, with $\tau \in [0,t]$,  and has not passed through the origin before that.  Consequently,  one has
\begin{align}
{\mathcal P}_{\rm r}(x_0,t)&=-\frac{\partial S_{\rm r}(x_0,t)}{\partial x_0}=\int_0^t {\rm d}\tau\,  S_{\rm r}(x_0,t-\tau)r e^{-r \tau}{\mathcal P}(x_0,\tau)+e^{-r t}{\mathcal P}(x_0,t).
\label{eq:reset-fpt-renewal}
\end{align}
Indeed,  trajectories that did not have any reset during the time interval $[0,t]$ (the probability of which, according to our earlier discussions,  is $\exp(-rt)$) give rise to the second term on the rhs of the second equality above.  On the other hand,  trajectories that at time $t$ have had last reset in the interval $[t-\tau-{\rm d}\tau,t-\tau]$ (the probability of which is $\exp(-r\tau) \,r{\rm d}\tau$), with $\tau \in [0,t]$,  and have not hit the origin before that give rise to the first term on the rhs of the second equality.
 
Using Eq.  (\ref{eq:PFPT-diffusion}) in Eq.  (\ref{eq:reset-fpt-renewal}),  taking Laplace transform of both sides,  and using $S_{\rm r}(x_0,t=0)=1$,  we finally get
\begin{align}
\widetilde{S}(x_0,s)=\frac{1-A}{r A+s},
\end{align}
where the quantity $A$ is defined as
\begin{align}
A \equiv A(s,r)= {\mathcal L}\left[\frac{x_0 e^{-r t -x_0^2/(4Dt)}}{\sqrt{4\pi Dt^3}}\right]=e^{-\sqrt{(x_0^2/D)(s+r)}}.
\end{align}
Here,  we have used ${\mathcal L}[t^{-3/2}\exp(-a/(4t))]=2\sqrt{\pi/a}\exp(-\sqrt{as})$ for ${\rm Re}(a)>0$.  
Consequently,  we obtain the MFPT in presence of resetting as \cite{Evans:2011}
\begin{align}
\langle t_f\rangle_{\rm r}=\frac{e^{x_0\sqrt{r/D}}-1}{r}.
\label{eq:MFPT-resetting}
\end{align}

\begin{figure}
\centering
\includegraphics[width=85mm]{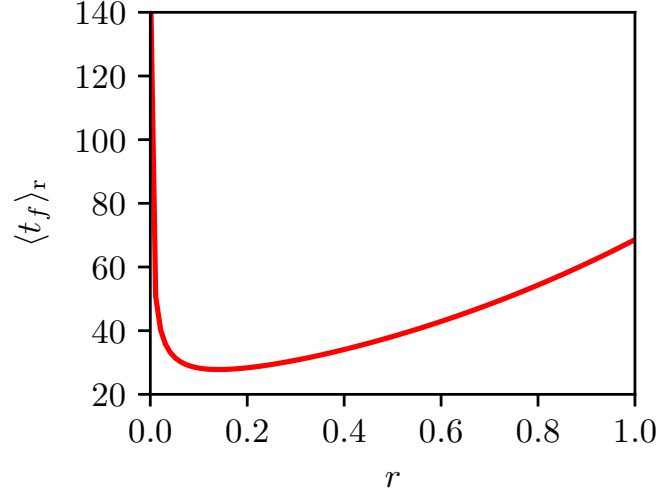}
\caption{The mean first-passage time in presence of resetting given by Eq.  (\ref{eq:MFPT-resetting} is shown as a function of $r$ for $x_0=3.0$ and $D=0.5$. One may observe the existence of a distinct minimum.}
\label{fig:MFPT-reset}
\end{figure}

We see from Eq.  (\ref{eq:MFPT-resetting}) that for fixed $x_0$ and $D$,  the MFPT in presence of resetting is finite for finite $r$,  so that resetting has a drastic consequence on rendering the MFPT of free diffusion finite.  Moreover, $\langle t_f \rangle_{\rm r}$ diverges in two extreme limits,  namely, as $r \to 0$ and as $r \to \infty$,  so it has to be that there exists a value $r^\star$ of $r$ at which $\langle t_f \rangle_{\rm r}$ as a function of $r$ has a minimum. Note that  the limit $r \to 0$ corresponds to free diffusion,  so that the  fact of a diverging $\langle t_f \rangle_{\rm r}$ as $r \to 0$ is consistent with what we observed above regarding the MFPT for free diffusion.  The divergence of MFPT as $r \to \infty$ may be understood on the basis of the fact that in this limit,  the particle resets so often that it does not effectively get a chance to move away from $x_0$ and hit the origin.  The quantity $r^\star$ satisfying ${\rm d}\langle t_f\rangle_{\rm r}/{\rm d}r|_{r=r^\star}=0$ yields the following transcendental equation \cite{Evans:2011}
\begin{align}
\frac{z^\star}{2}=1-e^{-z^\star};~~r^\star=(z^\star)^2D/x_0^2,
\label{eq:optimal-r}
\end{align}
which has a unique non-zero solution $z^\star=1.59362\ldots$.  It then follows that there is an optimal resetting rate $r^\star$,  such that the MFPT in presence of resetting attains its minimum as a function of $r$.  The variation of $\langle t_f\rangle_{\rm r}$ with $r$, as given by Eq.  (\ref{eq:MFPT-resetting}), is shown in Fig.  \ref{fig:MFPT-reset}, which clearly shows the existence of a minimum.

\begin{figure}
\centering
\includegraphics[width=85mm]{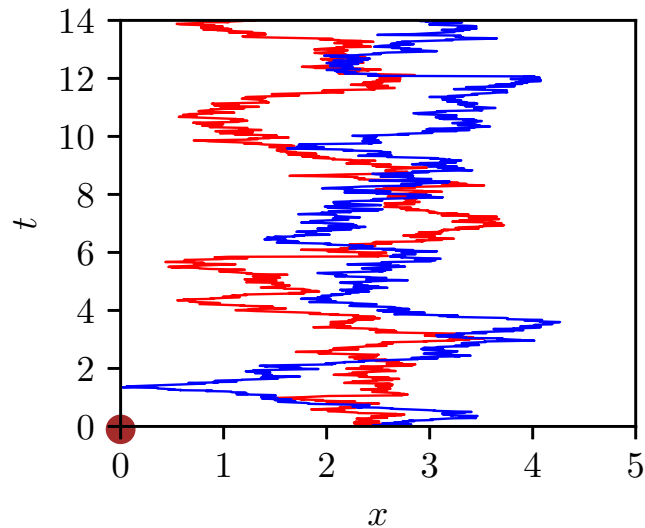}
\caption{The plot shows a stationary target (misplaced belonging) denoted by the brown filled circle and located at the origin,  and two space-time trajectories of a particle starting from $x_0=2.5$ and executing stochastic resets to $x_0$ at random times.  The parameter values are $D=0.5$ and $r=0.25$.  From the figure, we see that one of the trajectories has been able to detect the target, while the other one is yet to locate it.} 
\label{fig:search}
\end{figure}

\subsection{Relevance of resetting in search processes}
What has then a minimum MFPT got to do with search processes? Imagine having misplaced one of your favourite or essential belongings. It could be your car key or your mobile phone.  After getting into your office, you usually keep the key on your office table.  Having misplaced it,  a typical tendency to locate it is to begin searching in random directions for the key (well, you do not know beforehand which particular direction to search for, so random directions would be the best bet!) while starting from your office.  Having searched unsuccessfully for sometime, you realize that perhaps you did not search well enough in the region around the table (after all, you usually keep the key on the table itself),  so you return to the table,  and re-initiate your search.  You keep repeating the process until you locate the key, and of course the first time you locate it,  you stop searching any further.  Figure~\ref{fig:search} models this search strategy in one dimension: the brown filled circle denotes your misplaced belonging,  while searching in random directions for random times followed by re-initiation of the search process from your starting position may be modelled as free diffusion interspersed in time with stochastic resets to the initial location (the two space-time trajectories in the plot),  namely, what has been our model of study in this work.  Had there been no resets, that is,  if you had continued searching in random directions without ever coming back to your initial location,  we would arrive at the following conclusion based on our analysis of free diffusion pursued above. If you had repeated the process over and over again,  then, in some realizations of the process,  you would have located the brown circle in a finite time, while there would be happily a large number of other realizations in which you fail to locate it in finite time, so that the MFPT through the brown circle would diverge ! On the other hand,  introducing stochastic resetting into the dynamics would make the MFPT finite,  as our theoretical analysis of diffusion with stochastic resetting has revealed.  Not just that,  there exists an optimal value for the rate $r$ at which stochastic resets are executed such that the corresponding MFPT attains its minimum value.  Thus, stochastic resetting offers a very efficient way to locate your misplaced belonging.  Of course, for resetting to be efficient,  the quantity $x_0$,  the initial location of the searcher to which it repeatedly resets, has to be small enough. In other words,  the location $x_0$ has to be close enough to the target for resetting to do a good job. Indeed, as is evident from Eq.~(\ref{eq:MFPT-resetting}),  if $x_0$ is large, the MFPT would be very large, and then resetting would not help to locate the target in a finite time.

\section{Stochastic thermodynamics of resetting}
In this section, we review briefly thermodynamics of systems subject to stochastic resetting, following Refs.~\cite{Fuchs:2016} and~\cite{Busiello:2020},  and using basic principles extending thermodynamic laws to nonequilibrium systems laid down and reviewed in Ref.~\cite{Seifert:2012}. 
Consider first a Markov process in which the state space instead of being continuous, as in Eq.~(\ref{eq:eom}), is discrete.  Let us denote the latter by non-zero integers $i$.  The set of all possible states spans the state space of the system. Let $N$ be the total number of states accessible to the system. The dynamics of the system is dictated by a set of transition rates $W_{i \to j} \ge 0$ giving the probability per unit time for the system to make a transition from state $i$ to state $j$.   We consider these transition rates to be time-independent.  Define $p_i(t)$ to be the probability for the system to be in state $i$ at time $t$, with the normalization $\sum_{i=1}^N p_i(t)=1~\forall~t$.  These probabilities evolve in time following the so-called Master equation~\cite{Bertin-book}
\begin{align}
\frac{{\rm d}p_i(t)}{{\rm d}t}=\sum_{j=1}^N \left[W_{j \to i}~p_j(t) - W_{i \to j}~p_i(t)\right],
\label{eq:master-equation}
\end{align}
where the first term (respectively, the second term) on the right hand side summarizes all possible ways in which $p_i(t)$ increases (respectively, decreases) in an infinitesimal time interval $[t,t+{\rm d}t]$ due to transitions from all states $j$ to the state $i$ in question (respectively,  due to transitions from state $i$ to states $j$).  Note that the Master equation~(\ref{eq:master-equation}) conserves in time the normalization of the probability $p_i(t)$. Let us note that the Brownian motion~(\ref{eq:eom}) may be considered as the continuous-space limit of a discrete-space continuous-time random walker moving with equal probability between the nearest-neighbour sites of a one-dimensional lattice~\cite{Bertin-book}; in the latter case, the state space is constituted by the site indices, and we have $W_{i \to j}=W_{j \to i}$. 

Now, for a given pair of states $i$ and $j$,  consider the following situation~\cite{Busiello:2020}: (i) Both the forward and backward transitions are allowed (denote the corresponding transition rates by $W_{i\to j}\equiv w_{i \to j}>0$ and $W_{j \to i}\equiv w_{j \to i}>0$).  (ii) Either the forward or the backward transition is allowed but not both (i.e., either we have $W_{i \to j} \equiv y_{i \to j}>0$ and $W_{j \to i}\equiv y_{j \to i}=0$, or, $W_{j \to i} \equiv y_{j \to i}>0$ and $W_{i \to j}\equiv y_{i \to j}=0$).  The former dynamics is known as the bidirectional process, while the latter is referred to as the unidirectional process~\cite{Busiello:2020}. The reason we want to focus on this specific situation is that it models the resetting dynamics considered in this work. Indeed,  a given state $x(t)$ can in an infinitesimal time interval $[t,t+{\rm d}t]$ evolve to any other $x$ value allowed by the dynamics~(\ref{eq:eom}), just as any other $x$ value can evolve to the given value $x(t)$  in an infinitesimal time interval $[t,t+{\rm d}t]$ provided of course the dynamics~(\ref{eq:eom}) allows that.  On the contrary, the resetting dynamics is such that  in an infinitesimal time interval $[t,t+{\rm d}t]$, any $x$ value can evolve to the value $x_0$, but the reverse move from $x_0$ to any other $x$ value via the resetting dynamics is not allowed (it is allowed only via the Brownian motion dynamics~(\ref{eq:eom})). The Master equation~(\ref{eq:master-equation}) then reads~\cite{Busiello:2020} 
\begin{align}
\frac{{\rm d}p_i(t)}{{\rm d}t}=\sum_{j=1}^N \left[w_{j \to i} p_j(t) - w_{i \to j}p_i(t)\right]+\sum_{j=1}^N \left[y_{j \to i} p_j(t) - y_{i \to j}p_i(t)\right],
\label{eq:master-equation-1}
\end{align}
where the summation in the second term on the right hand side is considered restricted to those pair of states for which the reverse transitions are not allowed (i.e., either we have $y_{i \to j}>0$ and $y_{j \to i}=0$, or, $y_{j \to i}>0$ and $y_{i \to j}=0$). 

The (Shannon) entropy of the system at time $t$ is given by (we work in units in which the Boltzmann constant is set to unity):
\begin{align}
S_{\rm sys}(t)=-\sum_{i=1}^N p_i(t) \ln p_i(t),
\end{align}
where $p_i(t)$ is to be obtained by solving the Master equation~(\ref{eq:master-equation-1}) subject to a given initial condition $\{p_i(0)\}_{1 \le i \le N}$. Differentiating the above equation with respect to time, using Eq.~(\ref{eq:master-equation-1}), and the normalization $\sum_{i=1}^N p_i(t)=1$ yield straightforwardly the entropy production rate for the system~\cite{Busiello:2020}:
\begin{align}
\dot{S}_{\rm sys}(t)=\sum_{i,j=1}^N w_{j \to i}~p_j(t) \ln \left(\frac{p_j(t)}{p_i(t)}\right)+\sum_{i,j=1}^N y_{j \to i}~p_j(t) \ln \left(\frac{p_j(t)}{p_i(t)}\right),
\label{eq:entropy-equation-2}
\end{align}
where the dot denotes derivative with respect to time. 

Let us for the moment disregard any resetting-like transitions in the system.  Equation~(\ref{eq:entropy-equation-2}) then reads 
\begin{align}
\dot{S}_{\rm sys}(t)&=\sum_{i,j=1}^N w_{j \to i}~p_j(t) \ln \left(\frac{p_j(t)}{p_i(t)}\right)=\sum_{i,j=1}^N w_{j \to i}~p_j(t) \ln \left(\frac{w_{j \to i}~p_j(t)}{w_{i \to j}~p_i(t)}\right)-\dot{S}_{\rm env}(t);\label{eq:entropy-balance}\\
&\dot{S}_{\rm env}(t) \equiv \sum_{i,j=1}^N w_{j \to i}~p_j(t) \ln \left(\frac{w_{j\to i}}{w_{i\to j}}\right).
\end{align}
In the above,  we have defined $\dot{S}_{\rm env}$ as the environment entropy production,  related to the change in entropy of the environment corresponding to the system making transitions from every state $j$ to every state $i$. Here, the understanding is that it is the interaction with the environment that induces transitions between the states of the system, just as in the dynamics~(\ref{eq:eom}), the variable $x$ evolves in time due to the noise $\eta(t)$ arising because of the presence of the environment constituted by the surrounding fluid.  Equation~(\ref{eq:entropy-balance}) then yields the total entropy production rate as
\begin{align}
\dot{S}^{\rm tot}(t)&\equiv \dot{S}_{\rm sys}(t)+\dot{S}_{\rm env}(t)\nonumber \\&=\sum_{i,j=1}^N w_{j \to i}~p_j(t) \ln \left(\frac{w_{j \to i}~p_j(t)}{w_{i \to j}~p_i(t)}\right)\nonumber \\
&=\frac{1}{2}\sum_{i,j=1}^N \left(w_{j \to i}~p_j(t) - w_{i \to j}~p_i(t)\right) \ln \left(\frac{w_{j \to i}~p_j(t)}{w_{i \to j}~p_i(t)}\right) \ge 0.
\end{align}
That $\dot{S}^{\rm tot}(t) \ge 0$ may be understood by noting that $(x-y)\ln(x/y) \ge 0$ for any real, positive $x, y$ and that $f \ln f=0$ for $f=0$.
The above expression for $\dot{S}^{\rm tot}$ is the Schnakenberg's formula for the total entropy production of the system~\cite{Busiello:2020}. 

In presence of resetting, Eq.~(\ref{eq:entropy-equation-2}) then leads to
\begin{align}
&\dot{S}^{\rm tot}(t)=\dot{S}_{\rm sys}(t)+\dot{S}_{\rm env}(t)-\dot{S}_{\rm reset}(t); \label{eq:entropy-balance-reset}\\
&\dot{S}_{\rm reset}(t) \equiv \sum_{i,j=1}^N y_{j \to i}~p_j(t) \ln \left(\frac{p_j(t)}{p_i(t)}\right),
\end{align}
where we have defined $\dot{S}_{\rm reset}(t)$ as the entropy production rate due to unidirectional processes, i.e.,  resetting-like transitions.  For the case when the system resets to a state $i_0$ from any state $j$ with a resetting rate $r_j$, we have
\begin{align}
&\dot{S}_{\rm reset}(t) =\sum_{j=1}^N r_j~p_j(t) \ln \left(\frac{p_j(t)}{p_{i_0}(t)}\right)=\dot{S}_{\rm reset}^{\rm abs}(t)+\dot{S}_{\rm reset}^{\rm ins}(t); \label{eq:entropy-reset}\\
&\dot{S}_{\rm reset}^{\rm abs}(t) \equiv \sum_{j=1}^N r_j~p_j(t) \ln \left(p_j(t)\right),\\
&\dot{S}_{\rm reset}^{\rm ins}(t)\equiv -\ln (p_{i_0}(t))\sum_{j=1}^N r_j~p_j(t).
\end{align}
Here,  $\dot{S}_{\rm reset}^{\rm abs}(t)$ is the absorption entropy rate, corresponding to the change in the Shannon entropy of the system due to the probability flux out of every state $j$ owing to resetting. On the other hand, the quantity $\dot{S}_{\rm reset}^{\rm ins}(t)$ is the insertion entropy rate that depends on the probability at the resetting state $i_0$ and the probability flux out of every other state $j$~\cite{Fuchs:2016}.

In a stationary state,  all the quantities $\dot{S}_{\rm sys}$,  $\dot{S}_{\rm env}$, $\dot{S}^{\rm tot}$,  $\dot{S}_{\rm reset}$, etc are time independent, and moreover, $\dot{S}_{\rm sys}=0$ by construction.  Hence,  with $\dot{S}^{\rm tot} \ge 0$,  we get from Eq.~(\ref{eq:entropy-balance-reset}) that
\begin{align} 
\dot{S}^{\rm tot}=\dot{S}_{\rm env}-\dot{S}_{\rm reset} \ge 0.
\label{eq:entropy-balance-steady-state}
\end{align}
In an equilibrium stationary state,  which is obtained in the absence of resetting (see the paragraphs following Eq.~(\ref{eq:MSD-BM-reset})),  one has the condition of detailed balance,  $w_{j \to i}~p_j = w_{i \to j}~p_i$ for all pairs of states $i$ and $j$ with time-independent $p_i$'s.  Consequently, one has $\dot{S}^{\rm tot}=0$,  and Eq.~(\ref{eq:entropy-balance-steady-state}) implies that $\dot{S}_{\rm env} = 0$.  One may have time-independent $p_i$'s (i.e., a stationary state) with violation of detailed balance; this leads to a non-vanishing $\dot{S}^{\rm tot}$, which is thus a fingerprint of non-equilibrium stationary states.  Resetting induces such a state, and hence,  in a nonequilibrium stationary state induced by resetting,  one would have
\begin{align}
\dot{S}_{\rm env}-\dot{S}_{\rm reset} > 0,
\end{align} 
which is interpreted as the second law of thermodynamics in presence of resetting~\cite{Fuchs:2016}.
Considering the particular example of a discrete-space continuous-time random walker, for which one has $w_{i \to j}=w_{j \to i}$,  we have $\dot{S}_{\rm env}=0$, and hence that $\dot{S}_{\rm reset} < 0$. This result should also apply to the dynamics~(\ref{eq:eom}) in presence of resetting, as we will demonstrate in the following. 

Equation~(\ref{eq:entropy-reset}) when generalized to a Markov process with continuous state space reads~\cite{Fuchs:2016}
\begin{align}
\dot{S}_{\rm reset}(t) = \int {\rm d}x~r(x) P_{\rm r}(x,t) \ln \left(\frac{P_{\rm r}(x,t)}{P_{\rm r}(x_0,t)}\right).
\end{align}
For the case $r(x)=r$, a space-independent resetting rate, the situation we have considered in this work (see Eq.~(\ref{eq:resetting-rules})),  we get
\begin{align}
\dot{S}_{\rm reset}(t) = r \int {\rm d}x~P_{\rm r}(x,t) \ln \left(\frac{P_{\rm r}(x,t)}{P_{\rm r}(x_0,t)}\right).
\end{align}

In the nonequilibrium stationary state~(\ref{eq:steadystate}),  we get
\begin{align}
\dot{S}_{\rm reset}=-\frac{r}{2}\int_{-\infty}^\infty {\rm d}y~|y-y_0|e^{-|y-y_0|}=-r.
\end{align}
We thus see explicitly that we have $\dot{S}_{\rm reset} < 0$ in the nonequilibrium stationary state induced in the Brownian dynamics due to resetting.  That the resetting entropy decreases in time is a reflection of the fact that what resetting essentially does is to reduce the uncertainty in the particle position.  To wrap up this part,  we mention a few relevant recent work dealing with thermodynamics of resetting: Ref.~\cite{Pal:2017} that discusses the validity of the so-called integral fluctuation theorems in presence of stochastic resetting, Ref. ~\cite{Gupta:2020} that considers an overdamped Brownian particle in a potential well modulated through an external protocol and subject to stochastic resetting and studies the fluctuations of the work done on the system, Ref.~\cite{Pal:2021} that derives a thermodynamic relation for systems with unidirectional transitions including in particular a random walk subject to stochastic resetting.

In closing this section, we indicate how resetting may play a role in optimizing the efficiency of stochastic heat engines.  We first describe briefly the principle of working of a stochastic heat engine constituted by a Brownian particle of mass $m$, which is placed in a fluid medium in equilibrium at a given temperature, thereby modelling a heat bath or a thermal reservoir~\cite{engine}.  For simplicity, let us consider the motion of the particle in one dimension $x$.  The particle is subject to trapping due to an attractive force derived from a time-dependent potential $U(x(t);t)$ and is also acted upon by a time-dependent non-conservative force $F(x(t);t)$.  The position $x(t)$ and the velocity $v(t)={\rm d}x(t)/{\rm d}t$ of the particle evolve in time following the underdamped Langevin dynamics:
\begin{align}
m\frac{{\rm d}^2 x(t)}{{\rm d}t^2}=-\frac{\partial U(x(t);t)}{\partial x}+F(x(t);t)-\gamma \frac{{\rm d}x(t)}{{\rm d}t}+\xi(t),
\label{eq:eom1}
\end{align}
where $\gamma$ is the friction coefficient accounting for the friction that the particle experiences while moving in the surrounding medium,  while $\xi(t)$ models the random force that the surrounding fluid molecules impart on the particle.  One models $\xi(t)$ as a Gaussian, white noise with zero mean and correlations in time given by $\langle \xi(t) \xi(t') \rangle = 2\gamma  T\delta(t-t')$, where $T$ is the temperature of the fluid medium in units of the Boltzmann constant.  Note that in the absence of the potential $U(x(t);t)$ and the non-conservative force $F(x(t);t)$, when one considers the limit of large damping (the limit $\gamma/m \to \infty$ at a fixed and finite $m$),  the dynamics~(\ref{eq:eom1}) takes the form of the overdamped Langevin dynamics~(\ref{eq:eom}).  The dynamics~(\ref{eq:eom1}) is what is known to physicists as the Brownian motion, while mathematicians prefer to refer to the overdamped dynamics~(\ref{eq:eom}) as the Brownian motion.

In using the system~(\ref{eq:eom1}) as a heat engine, one allows the temperature of the surrounding fluid medium to switch between a higher temperature $T_h$ and a lower temperature $T_c$,  thus mimicking the hot and the cold thermal reservoir of an engine,  respectively.  A prominent example of an engine is the well-studied Carnot engine.  Specifically,  in our setup,  considering $U$ to be a harmonic potential,  one varies the stiffness of this potential in time from times $0$ to $t_1$ to implement the isothermal expansion at temperature $T_h$ and from times $t_1$ to $t_1+t_3$ to implement the isothermal compression at temperature $T_c$ of a Carnot cycle,  while the adiabatic expansion and compression stages of the latter may be implemented by performing an instantaneous change in the stiffness at time instants $t_1$ and $t_1+t_3$, with concomitant jump in temperature from $T_h$ to $T_c$ and from $T_c$ to $T_h$, respectively.  Heat transfer takes place between the bath and the particle during the isothermal stages and not during the adiabatic stages.  The aforementioned scenario defines a stochastic Carnot heat engine at the micro scale,  with reasons of stochasticity being enhanced thermal fluctuations prevalent in the dynamics at such scales~\cite{Seifert-engine}.

The heat transfer between the heat bath and the particle in time $t$ and computed along the trajectory of the particle from $x(0)$ to $x(t)$ is
given by
\begin{align}
Q(t)=\int_{x(0)}^{x(t)} \left(-\gamma \frac{{\rm d}x(t')}{{\rm d}t'}+\xi(t')\right)\circ {\rm d}x(t'),
\label{eq:heat}
\end{align}
where $\circ$ implies that the integral on the right hand side is to be evaluated in the Stratonovich sense~\cite{Gardiner}.  Similarly,  the work done on the particle in time $t$ equals
\begin{align}
W(t) = \int_0^t \frac{\partial U(x(t');t')}{\partial t'}\circ {\rm d}x(t')+\int_0^t F(x(t');t')\circ {\rm d}x(t').
\label{eq:work}
\end{align}
Note that both the heat and the work are stochastic functions of time that fluctuate between trajectories of the particle.  Next, one invokes stochastic definition of efficiency as given by the ratio of the stochastic work extracted in a cycle to the stochastic heat transferred $Q_h(t)$ from the hot bath to the
particle in a cycle,  as 
\begin{align}
\eta(t) = - \frac{W(t)}{Q_h(t)},
\end{align}
where $t$ denotes the total duration of the cycle.
On considering the ratio of average quantities, the corresponding efficiency $\overline{\eta}(t) = - \langle W(t) \rangle / \langle Q_h(t) \rangle$ is bounded on the upper side by the second law of thermodynamics by $\eta_C = 1-T_c/T_h$, with $\eta_C$ being the Carnot efficiency.  The challenge then is to engineer an optimal protocol for changing the stiffness of the potential $U$ so as to maximize the efficiency $\overline{\eta}(t)$.  Since the heat $Q_h$ and the work $W$ are functionals of the trajectories of the particle as generated by the dynamics~(\ref{eq:eom1}),  one may ask: is there a location $x(0)$ or $x(t)$ for which values of such quantities are optimised? In case it is so, does it then prove beneficial to implement repeated resetting of the position of the particle to such locations so as to achieve an enhanced efficiency of the engine? Considering resetting at exponential times (the set-up of Eq.~(\ref{eq:resetting-rules})),  an issue of importance would be:  is there an optimal resetting rate $r$ for which one obtains maximum efficiency? Can one work out the stochastic energetics of the engine in presence of resetting? In the wake of recent surge in activity in the field of both stochastic thermodynamics and stochastic resetting, it is tempting to speculate that study of such questions, either in the framework outlined above or suitable modifications thereof, may yield useful and insightful results on the issue of optimization of thermal machines at micro scales.  The perspectives presented here may serve as a genesis for future research in this direction.   

\section{Resetting of scaled and fractional Brownian motion}

\subsection{Scaled Brownian motion}
Until now, we have considered standard Brownian motion~(\ref{eq:eom}) in discussing the effects of stochastic resetting. We now discuss briefly the corresponding effects on the so-called scaled Brownian motion (SBM)~\cite{SBM-reset1,SBM-reset2}, which in appropriate limits reduces to the standard Brownian motion.  SBM is used in such contexts as fluorescence recovery following photobleaching~\cite{SBM1},  anomalous diffusion in biophysics~\cite{SBM2,SBM3},  granular gas of viscoelastic particles~\cite{SBM4}, etc.  In contrast to the standard Brownian motion~(\ref{eq:eom}) that involves a time-independent diffusion coefficient,  the SBM involves a diffusion coefficient that is time dependent and in fact which scales as a power-law in time. Thus, the corresponding dynamics is given by~\cite{SBM-reset1}
\begin{align}
\frac{{\rm d}x}{{\rm d}t}=\eta(t),
\label{eq:eom-SBM}
\end{align}
with $\eta(t)$ a Gaussian, white noise with properties
\begin{align}
\langle \eta(t)\rangle=0,~~\langle \eta(t)\eta(t')\rangle=2D(t)\delta(t-t');
\end{align}
here, we have
\begin{align}
D(t)=\alpha K_\alpha t^{\alpha-1};~~\alpha>0,
\end{align}
with $K_\alpha$ a constant.  Note that setting $\alpha=1$ and $K_1=D$ reduces the SBM to the standard BM~(\ref{eq:eom}).  In the case of the SBM, the position probability distribution may be shown to have the form~\cite{SBM-reset1}
\begin{align}
P(x,t)=\frac{1}{\sqrt{4\pi K_\alpha t^\alpha}}e^{-(x-x_0)^2/(4K_\alpha t^\alpha)},
\label{eq:xdistribution-SBM}
\end{align}
which as expected reduces for $\alpha=1$ to the result~(\ref{eq:xdistribution}). The MSD for the SBM reads~\cite{SBM-reset1}
\begin{align}
\langle (x-x_0)^2\rangle=2K_\alpha t^\alpha.
\label{eq:MSD-SBM}
\end{align}
The behavior of the MSD as a function of time depends explicitly on the value of $\alpha$.  Obviously,  $\alpha=1$, the case of the standard BM, results in diffusive behavior. 
For $0<\alpha<1$, one has a subdiffusive behavior, while the behavior is superdiffusive for $\alpha>1$.  For $\alpha=2$, one has ballistic growth of the MSD in time,  and $\alpha > 2$ results in superballistic behavior.  In the limit $\alpha \to 0$,  one has logarithmic time dependence of the MSD.

In discussing resetting of the SBM, two separate situations were considered: one in which only the position of the particle resets to its initial value (referred to as nonrenewal resetting,~\cite{SBM-reset1}), and other in which both the position and the diffusion coefficient reset to their respective initial values (referred to as renewal resetting,~\cite{SBM-reset2}).  Let us discuss the effects of resetting in the two cases.  In the case of nonrenewal resetting at exponentially-distributed random times (the set-up of Eq.~(\ref{eq:resetting-rules})), it was shown that in the long-time limit (more precisely, for times $t$ satisfying $t^{\alpha+1} \gg (x-x_0)^2/(K_\alpha r)$),  the position probability distribution in presence of resetting is given by~\cite{SBM-reset1}
\begin{align}
P_{\rm r}(x,t)&\simeq \frac{1}{2}\sqrt{\frac{r}{\alpha K_\alpha}}t^{(1-\alpha)/2}\exp\left(-\sqrt{\frac{r}{\alpha K_\alpha}}|x-x_0|t^{(1-\alpha)/2}\right).
\label{eq:reset-Pxt-SBM-nonrenewal}
\end{align}
The above equation implies a probability distribution that is non-Gaussian and also time-dependent (unless $\alpha=1$), with a cusp at the location of resetting $x=x_0$; 
For $\alpha=1$, the result yields the stationary-state~(\ref{eq:steadystate}) for the standard BM.  The MSD in presence of resetting is given in the limit of long times by~\cite{SBM-reset1}
\begin{align}
\langle (x-x_0)^2\rangle(t) \simeq \frac{2\alpha K_\alpha}{r}t^{\alpha-1}.
\label{eq:MSD-SBM-nonrenewal-reset}
\end{align}
We thus see that with respect to the result for the SBM in absence of resetting,  Eq.~(\ref{eq:MSD-SBM}), the MSD in presence of resetting has a time dependence characterized by an exponent that is smaller by unity.  For $\alpha=1$,  Eq.~(\ref{eq:MSD-SBM-nonrenewal-reset}) yields the stationary-state result~(\ref{eq:MSD-BM-reset}) for the standard BM.  For $\alpha=2$, when the SBM in absence of resetting shows ballistic motion,  the MSD in presence of resetting exhibits normal diffusive behavior. Superdiffusive SBM corresponding to $\alpha $ in the range $1<\alpha <2$ exhibits a subdiffusive behavior in presence of resetting. For $0 < \alpha <1$, when the SBM in absence of resetting shows a subdiffusive behavior, Eq.~(\ref{eq:MSD-SBM-nonrenewal-reset}) shows that the MSD in presence of resetting decays to zero as a power law,  implying thereby that the particle at long times remains in the close vicinity of the resetting location.  We thus see that in the case of nonrenewal resetting of the SBM at exponentially-distributed random times,  one ends up with a rather rich variety of behavior of the position probability distribution and the MSD depending on the value of $\alpha$. In the foregoing, we discussed the case of resetting at exponentially-distributed times; for discussion on the effects of resetting at power-law-distributed times, the reader is referred to Ref.~\cite{SBM-reset1}.   

We now turn to the case of renewal resetting of the SBM. In this case,  if there is a resetting at time instant $t_r$, the particle resets its position to $x_0$, and also the diffusion constant at a later time $t>t_r$ becomes $D(t)=\alpha K_\alpha (t-t_r)^{\alpha-1}$ (in the nonrenewal case discussed above,  the diffusion coefficient is by contrast given by $D(t)=\alpha K_\alpha t^{\alpha-1}$). The case has been studied in detail for both exponential and power-law resetting in Ref.~\cite{SBM-reset2}. Here, in the spirit of this review,  we discuss only the former scenario, referring the interested reader to Ref.~\cite{SBM-reset2} for discussions on power-law resetting.  For exponential resetting, it was shown that the position probability distribution relaxes at long times to a stationary state~\cite{SBM-reset2}
\begin{align}
P_{\rm r, ss}(x) &\simeq \frac{r\sqrt{2}}{\sqrt{\alpha(\alpha+1)}}\left(\frac{\alpha}{4K_\alpha r}\right)^{1/(\alpha+1)}|x-x_0|^{(1-\alpha)/(\alpha+1)}\nonumber \\
&\hskip10pt \times \exp\left[-\left(\frac{(x-x_0)^2r^\alpha}{4K_\alpha}\right)^{1/(\alpha+1)}\left(\alpha^{1/(\alpha+1)}+\alpha^{-\alpha/(\alpha+1)}\right)\right].
\label{eq:steadystate-SBM-renewal}
\end{align}
Not surprisingly,  for $\alpha=1$, the above equation reproduces correctly the result~(\ref{eq:steadystate}) for the standard BM.
Corresponding to the stationary state~(\ref{eq:steadystate-SBM-renewal}),  the MSD attains the value~\cite{SBM-reset2}
\begin{align}
\langle (x-x_0)^2\rangle_{\rm r,ss}=\frac{2K_\alpha}{r^\alpha}\Gamma(\alpha+1),
\label{eq:MSD-SBM-reset-renewal}
\end{align}
where $\Gamma(x)$ is the Gamma function. Again, for $\alpha=1$, one gets back the result~(\ref{eq:MSD-BM-reset}) for the standard BM. On the basis of the above discussions we thus see a drastic difference between the cases of nonrenewal and renewal resetting. While in the former, resetting fails to induce a stationary position distribution, see Eq. ~(\ref{eq:reset-Pxt-SBM-nonrenewal}), it does suffice to induce a stationary distribution in the renewal case,  see Eq.~(\ref{eq:steadystate-SBM-renewal}). Of course,  for $\alpha=1$, when the diffusion coefficient has no dependence on time, there is no difference between the two cases of resetting, and one has a stationary state.  In Ref.~\cite{SBM-reset2},  the MFPT has been investigated for the case of renewal resetting. Considering  a stationary target located at the origin and an SBM starting at $x_0>0$ and resetting to $x_0$ at exponentially-distributed random times, the MFPT through the target is given by~\cite{SBM-reset2}
\begin{align}
\langle t_f\rangle_{\rm r} \simeq \frac{1}{r}\left\{\sqrt{\frac{\alpha+1}{2\alpha}}\exp\left[r^{\alpha/(\alpha+1)}\left(\frac{x_0^2}{4K_\alpha}\right)^{1/(\alpha+1)}\left(\alpha^{1/(\alpha+1)}+\alpha^{-\alpha/(\alpha+1)}\right)\right]-1\right\}.
\label{eq:MFPT-resetting-SBM-renewal}
\end{align}
For $\alpha=1$, one recovers the result~(\ref{eq:MFPT-resetting}) for the standard BM.  As a function of $r$, the above MFPT is minimized at $r=r^\star$ satisfying~\cite{SBM-reset2}
\begin{align}
1-\sqrt{\frac{2\alpha}{\alpha+1}}\exp\left(-B(r^\star)^c\right)=Bc(r^\star)^c;~~c \equiv \frac{\alpha}{\alpha+1},~B \equiv \left(\frac{x_0^2}{4K_\alpha}\right)^{1/(\alpha+1)}\left(\alpha^{1/(\alpha+1)}+\alpha^{-\alpha/(\alpha+1)}\right). 
\end{align}
The above equation for $\alpha=1$ reduces correctly to the result~(\ref{eq:optimal-r}) for the standard BM.

\subsection{Fractional Brownian motion}
We now discuss the case of fractional Brownian motion (FBM),  another variant of the standard BM~(\ref{eq:eom}). In this case, the dynamics is given by~\cite{FBMref0,FBMref1}
\begin{align}
\frac{{\rm d}x}{{\rm d}t}=\eta_H(t),
\label{eq:eom-FBM}
\end{align}
with $\eta_H(t)$ a Gaussian noise with zero mean, which unlike the standard BM and the SBM is correlated in time:
\begin{align}
\langle \eta_H(t)\rangle=0,~~\langle \eta_H(t)\eta_H(t')\rangle \simeq K_{2H}2H(2H-1)|t-t'|^{2(H-1)},
\label{eq:memory}
\end{align}
where $H \in (0,1)$ is the so-called Hurst exponent.  The concept of FBM is invoked in discussions of subdiffusive dynamics~\cite{FBM1},  in investigating individual trajectories of fluorescently-labelled telomeres in the nucleus of living human cells~\cite{FBM2}, in discussing passive, thermally driven motion of micron-sized tracers in hydrogels of mucins, the main polymeric component of mucus~\cite{FBM3}, etc. 

In absence of resetting, the MSD of the FBM behaves as
\begin{align}
\langle (x-x_0)^2\rangle(t)=2K_{2H}t^{2H},
\label{eq:MSD-ensemble}
\end{align}
implying thereby normal diffusion for $H=1/2$,  subdiffusion for $0<H<1/2$ and superdiffusion for $1/2<H<1$.  Resetting of FBM has been considered under the so-called fully-renewal scheme~\cite{Wang},  whereby the memory of noise correlations given in Eq.~(\ref{eq:memory}) is completely erased at each resetting, and this is the case we consider here.  In presence of exponential resetting, the MSD has the above behavior only for short times, while for times of order $(1/r)[\Gamma(2H+1)]^{1/(2H)}$ shows saturation to a plateau (pl) value given by
\begin{align}
\langle (x-x_0)^2\rangle_{\rm pl} \approx \frac{2K_{2H}\Gamma(2H+1)}{r^{2H}}.
\label{eq:MSD-ensemble-long}
\end{align} 
Comparing the above result with the one for renewal resetting of the SBM at exponentially-distributed times, Eq.~(\ref{eq:MSD-SBM-reset-renewal}), we find that the value is the same as that in the SBM case with $\alpha=2H$. In the same manner, the FBM with exponential resetting has a stationary state characterized by a time-independent position probability distribution, which at intermediate-to-large displacements is given by the result~(\ref{eq:steadystate-SBM-renewal}) for the renewal resetting of the SBM, with the substitution $\alpha=2H$. 

While the definition~(\ref{eq:MSD-ensemble}) for the MSD involves averaging at a given time instant $t$ over ensemble of statistically-independent values of $x(t)$, it is interesting to ask for the behavior of the MSD when computed along a single trajectory of $x(t)$ as a function of time $t$, and enquire about ensemble-averaged versus time-averaged mean-squared displacements whose non-equivalence implies breaking of ergodicity of the underlying dynamics.  The single-trajectory-based averaging along the time series of particle position $x(t)$ defines the time-averaged-MSD (TAMSD) as
\begin{align}
\overline{\delta^2(\Delta)} \equiv \frac{1}{T-\Delta}\int_0^{T-\Delta} {\rm d}t~[x(t+\Delta)-x(t)]^2,
\end{align}
where $\Delta$ is the lag time and $T$ is the length of the time series. On averaging over $N$ statistically-independent TAMSD realizations, the mean TAMSD is obtained as 
\begin{align}
\langle \overline{\delta^2(\Delta)} \rangle= \frac{1}{N}\sum_{i=1}^N \overline{\delta_i^2(\Delta)},
\end{align}
where the angular brackets denote averaging over noise realizations. Reference~\cite{Wang} performed a detailed study of the behavior of the TAMSD versus MSD to conclude that for short times,  one has the MSD and the TAMSD behaving respectively as $t^{2H}$ and $\Delta^{2H}$ for $0 < H <1/2$ and as $t^{2H}$ and $\Delta^1$ for $1/2 < H <1$, so that time averaging and ensemble averaging do not coincide for $1/2 < H<1$, while they do coincide for $0<H<1/2$.  For long times, however, the two averages coincide for all values of $H$ and have the behavior as in Eq.~(\ref{eq:MSD-ensemble-long}).  While the two averages coincide for the FBM in the absence of resetting, the foregoing suggests that the resetting dynamics of originally ergodic FBM for superdiffusive $H>1/2$ exhibits weak ergodicity breaking at short times.  For a more detailed survey of the topic, the reader is referred to Ref.~\cite{Wang}.

\section{Stochastic resetting: A bird's eye view of recent work and  applications}  
In this section, we provide a flavor of recent work on the theme of stochastic resetting through a random selection of a few articles.  For a more exhaustive list,  the reader is referred to the review~\cite{Evans:2020}.  We will also discuss some applications of the concept of resetting to physical systems.

Reference~\cite{satyapaper1} addressed resetting of a diffusing particle for a space-dependent
resetting rate,  and resetting to a random position drawn from a resetting
distribution,  Ref.~\cite{satyapaper2} studied effects of partial absorption on first-passage time problems in the case of diffusion with stochastic resetting,  Ref.~\cite{satyapaper3} considered in the setting of first-passage time problems for a diffusive particle with stochastic resetting the issue of optimal search time when compared against that of an effective equilibrium Langevin process with the same stationary distribution, 
Ref.~\cite{satyapaper4} studied diffusion in arbitrary spatial dimension in presence of a resetting process.  Reference~\cite{satyapaper5} studied search process in one dimension by considering an immobile target and a searcher that undergoes a discrete-time jump process with successive jumps drawn independently from an arbitrary jump distribution. In Ref.~\cite{satyapaper6},  a simple random walk in one dimension was studied,  wherein at each time step the walker resets to the maximum of the already visited positions,  Ref.~\cite{satyapaper7} considered a continuous-space and continuous-time diffusion process under resetting to a position chosen from the dynamical trajectory in the past according to a memory kernel,  Ref.~\cite{satyapaper8} studied stochastic resetting in the context of the fractional Brownian motion. Quantum dynamics in presence of stochastic reset was addressed in Ref.~\cite{satyapaper9},  stochastic resetting in the case of a particle undergoing run and tumble dynamics in one dimension, whereby its velocity reverses stochastically,  was studied in Ref.~\cite{satyapaper10}, the issue of a stochastic process undergoing resetting in a manner that following each resetting,  there a random
refractory period during which the process is quiescent and remains
at the resetting position was studied in Ref.~\cite{satyapaper11}. Reference~\cite{satyapaper12} considered several lattice random walk models with stochastic resetting to previously visited sites that are shown to exhibit a phase transition between an anomalous diffusive regime and a localization regime in which diffusion is suppressed;  The transition is a result of a single impurity site at which the resetting rate is lower than on other sites, and around which the walker spontaneously localizes.  Close to criticality, the
localization length is shown to diverge with a critical exponent that falls in the same class as the self-consistent theory of Anderson localization of waves in random media.  The distribution of additive functions of Brownian motion subject to stochastic resets was investigated in Ref.~\cite{satyapaper13}, while the well-known Ising model with stochastic resetting was addressed in Ref.~\cite{satyapaper14}. Reference~\cite{satyapaper15} is an experimental exploration of the optimal mean time for a free diffusing Brownian particle to reach a target in presence of resetting, while Ref.~\cite{satyapaper16} analysed the non-equilibrium steady states and first-passage properties attained with a Brownian particle in an external confining potential that is switched on and off stochastically, and Ref.~\cite{satyapaper17} computed the mean perimeter and the mean area of the convex hull of a two-dimensional isotropic Brownian motion in the presence of resetting.

Reference~\cite{palpaper1} studied the stationary state attained with a Brownian particle diffusing in an arbitrary potential and subject to stochastic resetting, Ref.~\cite{palpaper2} considered a Brownian particle diffusing in presence of time-dependent stochastic resetting, whereby the rate of resetting is a function of the time elapsed since the last reset event, Ref.~\cite{palpaper3} is devoted to developing a general approach to treat theoretically first passage under stochastic resetting, while its extension to also include branching was considered in Ref.~\cite{palpaper4}. Reference~\cite{palpaper5} investigated the dynamics of a Brownian particle diffusing in a one-dimensional interval with absorbing end points, Ref.~\cite{palpaper6} studied local time for a Brownian particle in presence of stochastic resetting: Given a Brownian trajectory, the local time is the
time the trajectory spends in a vicinity of its initial position. A Landau-like theory to study phase transitions in resetting systems was developed in Ref.~\cite{palpaper7}, while stochastic resetting in the context of the many-particle system of symmetric exclusion process was studied in Ref.~\cite{palpaper8}. Reference~\cite{palpaper9} considered diffusion with stochastic resetting in which the diffusing particle resets to the resetting location with a finite speed.  Reference~\cite{palpaper10} took into account the fact that getting from place to place takes time, with places further away taking more time to be reached, in extending
the theory of stochastic resetting to account for this inherent spatio-temporal coupling, while Ref.~\cite{palpaper11} reported on experimental realization of colloidal particle undergoing diffusion and resetting via holographic optical tweezers, Ref.~\cite{palpaper12} proposed a method of resetting in the context of a Brownian particle, whereby non-instantaneous returns are facilitated by an external confining trap potential centered at the resetting location.  Reference~\cite{palpaper13} studied stochastic resetting in the context of the time that it takes to reach a stable equilibrium point in the basin of attraction of a dynamical system.  Reference~\cite{palpaper14} studied an overdamped Brownian particle subject to stochastic resetting in one dimension in which the particle undergoes a finite-time resetting process facilitated by an external linear potential,  while Ref.~\cite{palpaper15} investigated first-passage properties in the context of one-dimensional confined lattice random walks. 

Reference~\cite{sidpaper1} studied the case of stochastic search process in one, two, and three
dimensions in which $N$ diffusing searchers that all start at the same location search,  with each searcher also resetting to its starting point,  Refs.~\cite{sidpaper2,sidpaper3} considered first-passage resetting, whereby the resetting of a random walk to a fixed position gets triggered by the first-passage event of the walk itself.  Reference~\cite{boyerpaper1} studied the dynamics of predator–prey systems, whereby preys are confined to a region of space and predators move randomly according to a power-law dispersal kernel, and additionally, there is stochastic resetting of the predators to the prey patch,  while the dynamics of random walks on arbitrary networks with stochastic resetting to the initial position was analysed in Ref.~\cite{boyerpaper2},  and the case of resetting to multiple nodes was considered in Ref.~\cite{boyerpaper3}. Reference~\cite{boyerpaper4} investigated the effects of resetting on the reaction time between a Brownian particle and a stochastically-gated target.  Reference~\cite{paper0} studied stochastic multiplicative process with reset events,  Ref.~\cite{paper1} analyses large deviations of a ratio observable in discrete-time reset processes.  where the ratio has the form of a current divided by the number of reset steps, Ref.~\cite{paper2} studied diffusive motion in presence of stochastic resetting of a test particle in a two-dimensional comb structure consisting of a main backbone channel with continuously-distributed side branches,  Ref.~\cite{paper3} considered the Brownian motion with stochastic resetting of a particle in a bounded circular two-dimensional domain while searching for a stationary target on the boundary of the domain, Ref.~\cite{paper4} studied resetting in the context of the The Sisyphus random walk,  an infinite Markov chain whose
dynamics is such that at every clock tick,
the process can move rightward (or upward) one step or return to the initial state.  A unified renewal approach to the problem of random search for several targets under resetting was developed in Ref.~\cite{paper5}, while Ref.~\cite{paper6} considered a generalization of the basic set-up of stochastic resetting to the origin by a diffusing particle, whereby the diffusing particle may be only partially reset towards the trajectory origin or even overshoot the origin in a resetting step.  Reference~\cite{paper7} studied stochastic resetting in the framework of the monotonic continuous-time random walks with a constant drift. In Ref.~\cite{paper8},  for stochastic resetting of a random-walk process, a general perspective through derivation and analysis of mesoscopic (continuous-time random walk) equations for both jump and velocity models with stochastic resetting was offered.  Reference~\cite{paper9} explored within the context of resetting the possibility of observing an interesting dynamics, including phase transitions for the minimization of the MFPT,  for random walks with exponentially-distributed flights of constant speed,  Ref.~\cite{paper10} studied the effects of a stochastic resetting within the ambit of the exactly solvable one-dimensional coagulation–diffusion process,  Ref.~\cite{paper11} studied the effects of resetting on random processes that follow the so-called telegrapher’s equation.  Reference~\cite{paper12} explored first-passage properties resulting from an interplay of stochastic resetting with trapping due to an external potential, in the framework of a diffusing particle in a one-dimensional trapping potential, while Ref.~\cite{paper13} studied how active transport processes in living cells can be modelled by using the framework of a directed search process with stochastic resetting and delays.  

The concept of stochastic resetting has been invoked over the years in many different fields, from biology and ecology to computer science and psychology, to name a few.  Resetting finds applications in discussing search algorithms in computer science, e.g., in discussing  return to shallow points in a search tree by backtracking methods~\cite{app1},  and in addressing randomized search algorithms for hard combinatorial problems~\cite{app2}; in the field of psychology, e.g., to discuss pattern learning and recognition~\cite{app3}, and to optimize visual search~\cite{app4}; in the field of quantitative finance, e.g.,  in discussing reset options whereby the strike price of the option is reset periodically over the option's lifetime, to bring out-of-the-money options back to being at the money~\cite{app5}, and in deriving a valuation equation for a
European-style bear market warrant with a single
reset date~\cite{app6}; in biology, e.g.,  in considerations of a stochastic biophysical model for the motion of RNA polymerases during transcriptional pauses~\cite{app7}, and in the context of protein searching and recognition of targets on DNA~\cite{app8}; in ecology, e.g.,  in addressing relocation of animals to already visited places~\cite{app9}, and in movement ecology~\cite{app10}.  

\section{Summary and Conclusions}
Diffusion with stochastic resetting has been extensively studied in recent times,  and Ref.  \cite{Evans:2011} is considered a landmark contribution in recent times in the arena of nonequilibrium statistical physics,  which has really ushered in a new beginning in studies of stochastic processes.  While ours is a very brief review,  with the contributions of one of the authors of the current article being Refs.  \cite{ref2,ref1,ref3,ref4,ref5},  we refer the reader to the recent exhaustive review \cite{Evans:2020} to get a broader view of this exciting topic of research.  We hope that the current contribution will serve as an invitation to young minds to delve into the fascinating and exciting world of stochastic processes in general and stochastic resetting in particular.

\section*{Appendix: Discretization of Gaussian, white noise}
\label{app}
Here,  we discuss discrete-time representation of the noise $\eta(t)$ appearing in Eq.  (\ref{eq:eom}).  To this end,  we discretize time in
small steps of length $0< \Delta t \ll 1$,  so that the $i$-th time step is $t_i = i \Delta t$, with $i \in \mathbb{Z}$.  For an arbitrary function $f(t)$ of continuous time, one may write by using the definition of the Kronecker delta that the corresponding discrete-time representation is
\begin{align}
f_i \equiv f(t_i) &= \sum^{\infty}_{j=-\infty} ~f_j ~\delta_{i,j} = \sum^{\infty}_{j=-\infty} ~\Delta t ~f_j ~\frac{\delta_{i,j}}{\Delta t}.  
 \label{eq:app-delta-to-dirac-x1}
\end{align}
Then, in the limit $\Delta t \to 0$, identifying $\sum^{\infty}_{j=-\infty} \Delta t \equiv \int_{-\infty}^\infty {\rm d}t'$,
and with $f_j = f(t'), f_i  = f(t)$, we obtain from
Eq.  (\ref{eq:app-delta-to-dirac-x1}) that
\begin{align}
f(t) = \int_{-\infty}^{\infty}{\rm d}t'\, \bigg[ \lim_{\Delta t \to 0} \frac{\delta_{i,j}}{\Delta t} \bigg] f(t'). \label{eq:app-delta-to-dirac-x2}
\end{align}
From the definition of the Dirac delta function, one has $
f(t) = \int_{-\infty}^{\infty}{\rm d}t' \,\delta(t-t')~f(t')$.
Hence,  Eq. (\ref{eq:app-delta-to-dirac-x2}) implies that
\begin{align}
\delta(t-t') = \lim_{\Delta \tau \to 0} \frac{\delta_{i,j}}{\Delta t}.  
\label{eq:app-delta-to-dirac-x4}
\end{align}
On the basis of the above, we may write using Eq.  (\ref{eq:noise-properties}) the following properties for the noise in discrete times:
\begin{align}
\langle \eta_i \rangle = 0 \, , \quad \langle \eta_i \eta_j \rangle = \frac{2D}{\Delta t}
\delta_{i,j}\,; \quad i,j \in \mathbb{Z},
\end{align}
with the understanding that the discrete-time representation of $\eta(t)$ is $\eta_i \equiv \eta(t_i)$.  In particular, we have $\langle \eta(0) \rangle=0$ and $\langle \eta^2(0)\rangle=2D/\Delta t$,  the properties we use in the main text preceding Eq.  (\ref{eq:Q-DE}).

\section*{Conflict of Interest Statement}
The authors declare that the research was conducted in the absence of any commercial or financial relationships that could be construed as a potential conflict of interest.

\section*{Author Contributions}
All the authors contributed equally to the work.

\section*{Funding}
SG acknowledges support from the Science and Engineering Research Board (SERB),  Government of India under SERB-TARE Scheme Grant No. TAR/2018/000023, SERB-MATRICS Scheme Grant No. MTR/2019/000560, and SERB-CRG Scheme Grant No. CRG/2020/000596. 

\section*{Acknowledgements}
SG thanks M. Chase, A. Gambassi, L. Giuggioli, S. N. Majumdar,  A. Nagar,  \'{E}. Rold\'{a}n,  G. Schehr, and G. Tucci for fruitful collaborations and insightful discussions on the topic of stochastic resetting. SG thanks Debraj Das, Sayan Roy and Mrinal Sarkar for comments on the manuscript and for help with figures.  SG also thanks ICTP–Abdus Salam International Centre for Theoretical Physics, Trieste, Italy, for support under its Regular Associateship scheme. 

\bibliographystyle{frontiersinHLTH&FPHY}

\end{document}